\documentstyle[12pt]{article}
\begin{document}

\title{NS Open Strings with $B$ Field \\
and Their Interactions with NS Closed Strings
}

\author{
  Akira Kokado\thanks{E-mail: kokado@kobe-kiu.ac.jp},\  
  Gaku Konisi\thanks{E-mail: konisi@kwansei.ac.jp} \  
  and \ 
  Takesi Saito\thanks{E-mail: tsaito@yukawa.kyoto-u.ac.jp} \\
  \\
 {\small{\it{Kobe International University, Kobe 655-0004, Japan${}^{\ast }$}}} \\
 {\small{\it{Department of Physics, Kwansei Gakuin University, Sanda 669-1337, 
 Japan${}^{\dagger ,\ddagger }$}}}}

\date{\small{Feb. 2002}}
\maketitle

\begin{abstract}
We consider interactions of a NS open string with first important states of a NS closed string, i.e., a closed-string tachyon and a graviton, where both ends of the NS open string are attached on a D-brane, and a constant background $B$ field is lying along directions parallel to the D-brane world volume. Contrary to general expectations, there are no constraints on these vertex operators coming from the $B$ field. However, we point out that these vertex operators have singularities at both ends of the NS open string when external momenta take some values. These kinds of singularities essentially come from the Dirichlet boundary conditions along directions transverse to the D-brane world volume.
\end{abstract}

\setlength{\parindent}{1cm}
\newpage

\renewcommand{\thesection}{\Roman{section}.}
\renewcommand{\theequation}{\arabic{section}.\arabic{equation}}
\setcounter{equation}{0}

\section{Introduction}
\indent

Recently a physics of the open-string tachyon condensation draws much attention~\cite{rf:1}. There is a conjecture that after the tachyon condensation, unstable D-branes disappear and end up with a bunch of closed strings. In order to describe such phenomena, there have so far been published many papers~\cite{rf:2}. On the other hand, especially interesting is a model of open strings propagating in a constant antisymmetric $B$ field background. Previous studies show that this model is related to the noncommutativity of D-branes~\cite{rf:3}, and in the zero slope limit to noncommutative gauge theories~\cite{rf:4}. More about this model, there arises an observation that the tachyon condensation is simpler in the case when large noncommutativity is introduced on the D-branes~\cite{rf:5}. \\
\indent
   Taking into account of the above observation, it should be fundamentally important to consider interactions of open strings with closed strings, where both ends of open strings are attached on a D-brane and a constant $B$ field background is lying along directions parallel to the D-brane world volume. In this paper we would like to construct vertex operators of a Neveu-Schwarz open string interacting with first important states of a NS closed string, i.e., a closed-string tachyon and a graviton. NS tachyons are known to be important when we consider a space-time supersymmetry-breaking and the unstable D-brane. In the bosonic-string case we have already considered the same vertex operators~\cite{rf:6}. See also Refs. ~\cite{rf:7,rf:8,rf:9}. \\
\indent
  Contrary to general expectations, we find that there are no constraints on the vertex operators coming from the $B$ field. However, we find that these vertex operators have singularities (including branch points) at both ends of the NS open string when external momenta take some values. These kinds of singularities essentially come from the Dirichlet boundary conditions along the directions transverse to the D-brane world volume. \\
\indent
     In Sec.2 the action and boundary conditions of the NS open string are given when the constant $B$ field background is present. In Sec.3 we construct vertex operators that describe the emission of a NS closed-string tachyon and a graviton out of the NS open string. The final section is devoted to concluding remarks.

\section{NS open string in a constant-background $B$ field}
\indent

The action of the NS open string in a constant $B$ background is
\begin{equation}
\label{201}
 S=\frac{1}{2\pi \alpha '}\int  dZ d\bar{Z} (\eta _{\mu \nu } + B _{\mu \nu })
 \bar{D}X^\mu (Z,\bar{Z})DX^\nu  (Z,\bar{Z})
\end{equation}
where 
\[
 Z=(z,\theta ) \quad \bar{Z}=(\bar{z},\bar{\theta }),
\]
\begin{equation} 
\label{202}
 D=\frac{\partial }{\partial \theta } + \theta \frac{\partial }{\partial z}, \quad
 \bar{D}=\frac{\partial }{\partial \bar{\theta }} + \bar{\theta }\frac{\partial }{\partial \bar{z}},
\end{equation}
\[
 X^\mu (Z,\bar{Z})=x^\mu (z,\bar{z}) + i\theta \varphi _1^\mu (z,\bar{z})
 + i\bar{\theta }\varphi _2^\mu (z,\bar{z}) + i\theta \bar{\theta }F^\mu (z,\bar{z}),
\]
in the familiar notations. In the following we set $\alpha '=1$ for simplicity. In terms of the component fields the action (\ref{201}) is expressed as
\begin{equation}
\label{203}
 S = \frac{1}{2\pi }\int d^2z(\eta _{\mu \nu } + B _{\mu \nu })
 (\bar{\partial }x^\mu \partial x^\nu 
 - \bar{\partial }\varphi _1^{\ \mu } \varphi _1^{\ \nu } 
 + \varphi _2^{\ \mu } \partial \varphi _2^{\ \nu }).
\end{equation}
Here we have eliminated the auxiliary field $F^\mu (z,\bar{z})$  by using its equation of motion, 
and $\partial =\frac{\partial }{\partial z}$, $\bar{\partial }=\frac{\partial }{\partial \bar{z}}$. \\ 
\indent

     The constant-background field $B_{\mu \nu }$ is assumed to be lying on a D-brane world volume, which extends along ($x^0$, $x^1$, $\ldots$, $x^p$)-directions. Both ends of the open string 
 $X^\mu (Z,\bar{Z})$  are attached on the D-brane. The boundary conditions on the string coordinates are imposed as follows:
\[
 \eta _{\mu \nu }(DX^\nu - \bar{D}X^\nu)+B_{\mu \nu }(DX^\nu + \bar{D}X^\nu)|_{\sigma =0,\pi ; \theta=\bar{\theta  }}=0, \quad (\mu ,\nu = 0,1,\ldots ,p)
\]
\begin{equation}
\label{204}
 DX^i + \bar{D}X^i|_{\sigma =0,\pi ; \theta=\bar{\theta  }}=0, \quad (i=p+1,\ldots ,9)
\end{equation}
where the latter is the Dirichlet one along directions transverse to the D-brane, whereas the former is a mixed one of Neumann and Dirichlet along directions parallel to the brane world volume including time. \\
\indent 
     These conditions are reduced to, for the bosonic components,
\[
  \eta _{\mu \nu }(\partial - \bar{\partial })x^\nu + 
  B_{\mu \nu }(\partial + \bar{\partial })x^\nu|_{\sigma =0,\pi}=0, \quad (\mu ,\nu = 0,1,\ldots ,p), 
\]
\begin{equation}
\label{205}
(\partial + \bar{\partial })x^i|_{\sigma =0,\pi}=0, \quad (i = p+1,\ldots ,9),
\end{equation}
and for fermionic components 
\[ 
  \eta _{\mu \nu }(\varphi _1^{\ \nu } - \varphi _2^{\ \nu }) + 
  B_{\mu \nu }(\varphi _1^{\ \nu } + \varphi _2^{\ \nu })|_{\sigma =0,\pi}=0, 
  \quad (\mu ,\nu = 0,1,\ldots ,p)
\]
\begin{equation}
\label{206}
 \varphi _1^{\ i} + \varphi _2^{\ i}|_{\sigma =0,\pi}=0. \quad (i = p+1,\ldots ,9)
\end{equation}
\indent
 The mode expansions along directions, $\mu = 0,1,\ldots ,p$, parallel to the D-brane are then given by, with a notation $B\equiv \tanh \beta $,
\[
 X(Z,\bar{Z}) = \frac{1}{\sqrt{2}}[e^{-\beta }Y(Z) + e^{\beta }Y(\bar{Z})].
\]
\begin{equation}
\label{207}
 Y(Z) = y(z) + i\theta \varphi (z),
\end{equation}
where
\[
 y(z) = q + \frac{1}{2}\pi \tanh \beta \cdot \alpha _0 + i\alpha _0\log z 
 + i\sum _{n\neq 0}\frac{1}{n}\alpha _nz^{-n}, 
\]
\begin{equation}
\label{208}
 \varphi (z) = \sum _{r\in Z+1/2}b_{r}z^{-r-1/2},
\end{equation}
and
\begin{equation}
\label{209}
 \varphi _1(z) = \varphi (z), \quad \varphi _2(\bar{z}) = \varphi (\bar{z}).
\end{equation}
The mode expansions along directions, $i=p+1,\ldots ,9$, transverse to the D-brane are, by setting $\beta =0$, 
\begin{equation}
\label{210}
 X(Z,\bar{Z}) = c + \frac{1}{\sqrt{2}}[Y(Z) - Y(\bar{Z})] 
\end{equation}
where $c$ is an arbitrary real constant. We find that the commutation relations among the oscillators are given by
\[
 [q, q]=0,
\] 
\begin{equation}
\label{211}
 [q, \alpha _n]=i\delta _{n,0},
\end{equation}
\[
 [\alpha _m, \alpha _n] = m\delta _{m+n,0},
\]
for all components, $\mu =0,1,\ldots ,9$.

\section{Vertex operators}
\setcounter{equation}{0}
\subsection{Closed-string tachyon vertex operator}
\indent

First we construct a vertex operator that describes the emission of a ground-state tachyon of a NS closed string out of a NS open string, when a constant background B field is lying along directions parallel to a D-brane world volume. \\
\indent
Let us define a contraction by
\begin{equation}
\label{301}
 \langle \bullet \bullet \bullet \rangle  = \bullet \bullet \bullet - :\bullet \bullet \bullet :. 
\end{equation}
where $:\bullet \bullet \bullet :$  denotes normal ordering. For the components $\mu =0,\ldots ,p$  we have
\[
 \langle Y_1Y_2\rangle  = \langle Y(Z_1)Y(Z_2)\rangle  = -\log Z_{12} - \frac{1}{2}i\pi \tanh \beta ,
\]
\begin{equation}
\label{302}
 \langle (DY)_1Y_2\rangle  = -\theta _{12}Z_{12}^{\ \ -1},
\end{equation} 
\[
 \langle (\partial Y)_1Y_2\rangle  = -Z_{12}^{\ \ -1},
\]
where
\[
 Z_{12} = z_1 - z_2 - \theta _1 \theta _2,
\]
\begin{equation}
\label{303}
 \theta _{12} = \theta _1 - \theta _2. 
\end{equation}
For the other components $i=p+1,\ldots ,9$ we have the same formulas as above but with $\beta =0$. \\
\indent
Let us put
\begin{equation}
\label{304}
 P=e^{-\beta }\Lambda _{//} + \Lambda _{\bot }, \qquad \bar{P}=e^\beta \Lambda _{//} - \Lambda _{\bot }, 
\end{equation}
where $\Lambda _{//} (\Lambda _{\bot })$ is a projection operator into directions parallel (transverse) to the D-brane.  Then we have
\begin{equation}
\label{305}
 X(Z,\bar{Z})=X(Z) + \bar{X}(\bar{Z}), 
\end{equation}
where
\begin{equation}
\label{306}
  X(Z) = \frac{1}{\sqrt{2}}PY(Z), \qquad \bar{X}(\bar{Z})= \frac{1}{\sqrt{2}}\bar{P}Y(\bar{Z}).
\end{equation}
\indent
Fundamental contractions are
\begin{equation}
\label{307}
 \langle PY_{1} PY_{2}\rangle = - \log Z_{12},
\end{equation} 
\begin{equation}
\label{308}
 \langle PY_{1} \bar{P}Y_{\bar{2}}\rangle = - Q \log Z_{1\bar{2}}, 
\end{equation}
where
\begin{equation}
\label{309}
 Q = e^{-2\beta }\Lambda _{//} - \Lambda_{\bot }.
\end{equation}
From Eqs. (\ref{307}) and (\ref{308}) it follows that
\[
 \langle (DX)_{1}X(Z_{2})\rangle = -\frac{1}{2}\theta _{12}Z_{12}^{\ \ -1},
\]
\begin{equation}
\label{310}
 \langle (\partial X)_{1}X(Z_{2})\rangle = -\frac{1}{2}Z_{12}^{\ \ -1},
\end{equation}
\[
 \langle (DX)_{1}\bar{X}(\bar{Z}_{2})\rangle = -\frac{1}{2}Q\theta _{1\bar{2}}Z_{1\bar{2}}^{\ \ -1},
\]
\begin{equation}
\label{311}
 \langle (\partial X)_{1}\bar{X}(\bar{Z}_{2})\rangle = -\frac{1}{2}QZ_{1\bar{2}}^{\ \ -1}
\end{equation}
\indent
Now, let us assume a vertex operator to be
\begin{equation}
\label{312}
 V(X(Z,\bar{Z}) = U(Z)\bar{U}(\bar{Z}),
\end{equation}
where
\[
U(Z) = :\exp\{ik\cdot X(Z)\}:,
\]
\begin{equation}
\label{313}
\bar{U}(\bar{Z}) = :\exp\{ik\cdot \bar{X}(\bar{Z})\}:.
\end{equation}
In the following we see that this is the right vertex operator for the NS closed-string tachyon. The super Virasoro operator is given by
\begin{equation}
\label{314}
 L(F) = \frac{1}{2\pi i}\oint dZ F(Z) T(Z),
\end{equation}
where the energy-momentum tensor is 
\begin{equation}
\label{315}
 T(Z) = :\partial X\cdot DX: = \frac{1}{2}:\partial Y\cdot DY:.
\end{equation}
We calculate an operator product expansion of $T(Z_{1})\bar{U}(\bar{Z}_{2})\equiv T_{1}\bar{U}_{2}$.  By using Eqs. (\ref{311}) its contraction is given as
\begin{eqnarray}
\label{316}
 \langle T_{1}\bar{U}_{2}\rangle &=& \langle :(\partial X)_{1}\cdot (DX)_{1}:\bar{U}_{2}\rangle  \nonumber \\
  &=& :(\partial X)_{1}\cdot \langle (DX)_{1}\bar{U}_{2}\rangle + (DX)_{1}\cdot \langle (\partial Y)_{1}\bar{U}_{2}\rangle \nonumber \\
 && + \langle (\partial X)_{1}^{\ \bullet }\cdot (DX)_{1}^{\ \bullet \bullet }\bar{U}_{2}^{\bullet ,\bullet \bullet }\rangle : 
 \nonumber \\
  &=& :(\partial X)_{1}\langle (DX)_{1}\bar{X}_{2}\rangle \cdot \partial _{\bar{X}}\bar{U}_{2} 
 + (DX)_{1}\cdot \langle (\partial X)_{1}\bar{X}_{2}\rangle \cdot \partial _{\bar{X}}\bar{U}_{2}  \nonumber \\
 && + \langle (\partial X^{\mu })_{1}\bar{X}_{2}^{\ \lambda }\rangle \langle (DX_{\mu })_{1}\bar{X}_{2}^{\ \rho }\rangle ik_{\lambda }
 ik_{\rho }\bar{U}_{2}: \nonumber \\
  &=& :-\frac{1}{2}(\partial X)_{1}\cdot Q\theta _{1\bar{2}}Z_{1\bar{2}}^{\ \ -1}
 \cdot \partial _{\bar{X}}\bar{U}_{2} 
 -\frac{1}{2}(DX)_{1}\cdot QZ_{1\bar{2}}^{\ \ -1}\cdot \partial _{\bar{X}}\bar{U}_{2} \nonumber \\
 && -\frac{1}{4}Q^{\rho \sigma }Q_{\rho }^{\ \lambda }\theta _{1\bar{2}}Z_{1\bar{2}}^{\ \ -2}
 k_{\sigma }k_{\lambda }\bar{U}_{2}:. 
\end{eqnarray}
\indent
 Note that
\[
 (\partial X_{\rho })_{1}Q^{\rho \mu }|_{Z_{1}=\bar{Z}_{2}} = (\bar{\partial }\bar{X}^{\mu })_{2} \qquad etc,
\]
\begin{equation}
\label{317}
 Q^{\rho \sigma }Q_{\rho }^{\ \lambda } = \eta ^{\sigma \lambda }.
\end{equation}
The operator product expansion of $T_1\bar{U}_2$ is then substituted into the standard formula for the commutator to yield
\begin{eqnarray}
\label{318}
 [L(F), \bar{U}(\bar{Z}_2)] &=& \frac{1}{2\pi i}\oint dZ_1 F(Z_1)T(Z_1)\bar{U}(\bar{Z}_2) \nonumber \\
  &=& -\frac{1}{2}F_{\bar{2}}(\bar{\partial }\bar{X}^{\mu })_2(\bar{\partial _{\mu }}\bar{U})_2 
 - \frac{1}{2}\{\bar{D}(F(\bar{Z})\bar{D}\bar{X}^{\mu })\}_2(\bar{\partial }_{\mu }\bar{U})_2
 - \frac{1}{4} k^2(\partial F)_{\bar{2}}\bar{U}_{2} \nonumber \\
  &=& -\frac{1}{2} F(\bar{Z})\bar{\partial }\bar{U} - \frac{1}{2}\bar{D}\{F(\bar{Z})\bar{D}\bar{U}\}
 - \frac{1}{4} k^2\{\bar{\partial }F(\bar{Z})\}\bar{U}|_{\bar{Z}=\bar{Z}_2} 
\end{eqnarray}
If $k^2=2$, the right-hand side becomes a sum of total derivatives:
\begin{equation}
\label{319}
 [L(F), \bar{U}(\bar{Z})] = -\frac{1}{2} \bar{\partial }\{F(\bar{Z})\bar{U}(\bar{Z})\}
 -\frac{1}{2} \bar{D}\{F(\bar{Z})\bar{D}\bar{U}(\bar{Z})\}.
\end{equation}
In the same way we have, using Eqs. (\ref{310}),
\begin{equation}
\label{320}
 [L(F), U(Z)] = -\frac{1}{2} \partial \{F(Z)U(Z)\}
  -\frac{1}{2} D\{F(Z)DU(Z)\}.
\end{equation}
\indent
 According to the formula
\begin{equation}
\label{321}
 [L(F), V] = [L(F), U\bar{U}] = [L(F), U]\bar{U} + U[L(F), \bar{U}],
\end{equation}
we finally get a desired commutator
\begin{equation}
\label{322}
 [L(F), V] = -\frac{1}{2}[\partial (FV) + \bar{\partial }(F(\bar{Z})V) + D(FDV) + \bar{D}(F(\bar{Z})\bar{D}V)].
\end{equation}
Here we have set the constraint
\begin{equation}
\label{323}
 k^2 = 2.
\end{equation}
Since the right-hand side of Eq. (\ref{322}) consists of total derivatives, the action of the vertex operator 
$V=U\bar{U}$  with the constraint (\ref{323}) turns out to be superconformally invariant.  \\
\indent
The mass of the NS open-string tachyon is given by $M_{\textrm{open}}^{\ \ \ \ \ \ 2} = -1/(2\alpha ')$, whereas that of the NS closed-string tachyon is $M_{\textrm{closed}}^{\ \ \ \ \ \ \ 2} = -2/\alpha '$.  Since the external momentum in our vertex operator is subject to the constraint $\alpha 'k^2=2=-\alpha 'M_{\textrm{closed}}^{\ \ \ \ \ \ \ 2}$, the external field corresponds to the NS closed-string tachyon. This shows that our vertex operator (\ref{312}) with the constraint (\ref{323}) describes the emission of the NS closed-string tachyon out of the NS open string. There are no constraints coming from the background $B$ field. \\

\subsection{Gravition vertex operator}
\indent
Next we construct a graviton vertex operator that describes the emission of a graviton as one of NS closed-string states out of the NS open string with the same background $B$ field. \\
\indent
Let us assume the graviton vertex operator to be
\[
 V_g=\epsilon _{\mu \nu }:DX^{\mu }U::\bar{D}X^{\mu }\bar{U}:,
\]
\begin{equation}
\label{324}
 U(Z) = :\exp\{ik\cdot X(Z)\}:,
\end{equation}
\[
 U(Z) = :\exp\{ik\cdot \bar{X}(\bar{Z})\}:,
\]
where the polarization tensor $\epsilon _{\mu \nu }(k)$ satisfies
\begin{equation}
\label{325}
 \epsilon _{\mu \nu }(k) = \epsilon _{\nu \mu }(k), \qquad \epsilon _{\mu \nu }(k)k^\nu = 0,
 \qquad \eta ^{\mu \nu }\epsilon _{\mu \nu }(k) = 0.
\end{equation}
Relevant contractions are added to Eqs. (\ref{310}) and (\ref{311}):
\[
 \langle (DX)_1(DX)_2\rangle = -\frac{1}{2}Z_{12}^{\ \ -1},
\]
\begin{equation}
\label{326}
 \langle (\partial X)_1(DX)_2\rangle  = \frac{1}{2}\theta _{12}Z_{12}^{\ \ -2},
\end{equation}
\[
 \langle (DX)_1(\bar{D}\bar{X})_2\rangle = -\frac{1}{2}QZ_{1\bar{2}}^{\ \ -1},
\]
\begin{equation}
\label{327}
 \langle (\partial X)_1(\bar{D}\bar{X})_2\rangle = \frac{1}{2}Q\theta _{1\bar{2}}Z_{1\bar{2}}^{\ \ -2}\bar{U}(\bar{Z}).
\end{equation}
Then contractions including $\bar{U}(\bar{Z})$ are calculated as
\[
 \langle (\partial X^\rho )_1\bar{U}_2\rangle  
 = \langle (\partial X^\rho )_1\bar{X}^\sigma (\bar{Z}_2)\rangle \bar{\partial }_\sigma \bar{U}_2
 = -\frac{1}{2}Q^{\rho \sigma }Z_{1\bar{2}}^{\ \ -1}\bar{\partial }_\sigma \bar{U}_2,
\]
\begin{equation}
\label{328}
 \langle (DX^\rho )_1\bar{U}_2\rangle = \langle (DX^\rho )_1\bar{X}^\sigma (\bar{Z}_2)\rangle \bar{\partial }_\sigma \bar{U}_2
 = -\frac{1}{2}Q^{\rho \sigma }\theta _{1\bar{2}}Z_{1\bar{2}}^{\ \ -1}\bar{\partial }_\sigma \bar{U}_2,
\end{equation}
to give
\begin{eqnarray}
\label{329}
 \langle T_1&:&(\bar{D}X^\mu \bar{U})_2:\rangle = \langle :\partial X_\rho DX^\rho )::(\bar{D}X^\mu \bar{U})_2:\rangle \nonumber \\
  &=&:(\partial X_\rho )_1\langle (DX^\rho )_1(\bar{D}X^\mu )_2\rangle \bar{U}_2: 
 + :(DX_\rho )_1\langle (\partial X^\rho )_1(\bar{D}X^\mu )_2\rangle \bar{U}_2: \nonumber \\
  &&+ :(DX_\rho )_1(\bar{D}X^\mu )_2\langle (\partial X^\rho )_1\bar{U}_2\rangle : 
 + :(\partial X_\rho )_1\langle (DX^\rho )_1\bar{U})_2\rangle (\bar{D}X^\mu )_2: \nonumber \\
  &&+ :(\partial X_\rho )_1^{\ \bullet }(DX^\rho )_1^{\ \bullet \bullet }(\bar{D}X^\mu )_2\bar{U}_2^{\ \bullet, \bullet \bullet }: 
 + (\partial X_\rho )_1^{\ \bullet }(DX^\rho )_1^{\ \bullet \bullet }(\bar{D}X^\mu )_2^{\ \bullet \bullet }\bar{U}_2^{\ \bullet } \nonumber \\
  &&+ (\partial X_\rho )_1^{\ \bullet }(DX^\rho )_1^{\ \bullet \bullet }(\bar{D}X^\mu )_2^{\ \bullet }\bar{U}_2^{\ \bullet \bullet } \nonumber \\
  &=& :-\frac{1}{2}(\partial X_\rho )_1Q^{\rho \mu }Z_{1\bar{2}}^{\ \ -1}\bar{U}_2 
 + \frac{1}{2}(DX_\rho )_1Q^{\rho \mu }\theta _{1\bar{2}}Z_{1\bar{2}}^{\ \ -2}\bar{U}_2 
 \nonumber \\
 && - \frac{1}{2}(DX_\rho )_1(\bar{D}X^{\mu })_2Q^{\rho \sigma }Z_{1\bar{2}}^{\ \ -1}\bar{\partial }_{\sigma }\bar{U}_2 \nonumber \\
  && - \frac{1}{2}(\partial X_\rho )_1Q^{\rho \sigma }\theta _{1\bar{2}}Z_{1\bar{2}}^{\ \ -1}\bar{\partial }_{\sigma }\bar{U}_2(\bar{D}X^{\mu })_2: \nonumber \\
 && + \frac{1}{4}Q^{\rho \sigma }Q_{\rho }^{\lambda }\theta _{1\bar{2}}Z_{1\bar{2}}^{\ \ -2}\bar{\partial }_{\sigma }\bar{\partial }_{\lambda }\bar{U}_2
 + \frac{1}{4}Q^{\rho \sigma }Q_{\rho }^{\ \mu }Z_{1\bar{2}}^{\ \ -2}\bar{\partial }_{\sigma }\bar{U}_2 
\end{eqnarray}
Noting Eqs. (\ref{317}), we have
\begin{eqnarray}
\label{330}
 [L(F), :(\bar{D}\bar{X}^{\mu }(\bar{Z})\bar{U}(\bar{Z}):] 
  = :-\frac{1}{2}\bar{D}\{\bar{D}(\bar{F}\bar{D}X^{\mu }\bar{U})+\bar{F}\bar{D}(\bar{D}X^{\mu }\bar{U})\} \nonumber \\
  -\frac{1}{4}k^2\bar{\partial }\bar{F}\bar{D}X^{\mu }\bar{U} 
  +\frac{1}{4}(\bar{\partial }\bar{D}\bar{F})\bar{\partial }^{\mu }\bar{U}:.
\end{eqnarray}
The last term vanishes owing to the constraints (\ref{325}) when multiplied by $\epsilon _{\mu \nu }$. If we set a constraint
\begin{equation}
\label{331}
 k^2 = 0,
\end{equation}
the right-hand side becomes a sum of total derivatives. In quite the same way we have, using Eqs. (\ref{326}),
\begin{eqnarray}
\label{332}
 [L(F), :DX^{\mu }(Z)U(Z):]
 = :-\frac{1}{2}D\{D(FDX^{\mu }U) + FD(DX^{\mu }U)\} \nonumber \\ - \frac{1}{4}k^2\partial FDX^{\mu }U 
 + \frac{1}{4}(\partial DF)\partial ^{\mu }U:.
\end{eqnarray}
Neglecting the last two terms and using the formula (\ref{321}), we finally find
\begin{equation}
\label{333}
 [L(F), V_g] = -\frac{1}{2}D\{D(FV_g)+FDV_g\} -\frac{1}{2}\bar{D}\{\bar{D}(\bar{F}V_g)+\bar{F}\bar{D}V_g\}.
\end{equation}
Since the right-hand side of Eq. (\ref{333}) is composed of total derivatives, we see that Eq. (\ref{324}) is the right graviton vertex operator. Again there are no constraints coming from the $B$ field.

\section{Concluding remarks}
\setcounter{equation}{0}
\indent 

 We have constructed most important vertex operators that describe the emissions of a ground-state tachyon and a massless-state graviton of a NS closed string out of a noncommutative NS open string, where a constant $B$ field background is lying along directions parallel to a D-brane world volume.  Contrary to general expectations, we have seen that there are no constraints coming from the $B$ field. \\
\indent
 Our vertex operator (\ref{312}) as well as (\ref{324}) is written as a product of chiral-vertex operators. They can be rewritten in totally normal-ordered forms, up to multiplicative constant factors, that is, for (\ref{312})
\begin{equation}
\label{401}
 V = U\bar{U} = \tilde{Z}^{\gamma (k)}:\exp\{ik\cdot X(Z, \bar{Z})\}:
\end{equation}
with
\begin{equation}
\label{402}
 \tilde{Z} \equiv z - \bar{z} - \theta \bar{\theta },  
\end{equation}
\begin{equation}
\label{403}
 \gamma (k) \equiv \frac{1}{2}(k_{//}\cdot \cosh 2\beta \cdot k_{//} - k_{\bot }^{\ 2}),  
\end{equation}
where $k_{//} (k_{\bot })$ stands for a momentum parallel (transverse) to the D-brane. One can see that this factor $\tilde{Z}^{\gamma (k)}$ in (\ref{401}) diverges at both end points, $z=\bar{z}$, $\theta =\bar{\theta }$,  of the NS open string, when $\gamma (k)<0$. More generally, it has a singular (branch) point at $\tilde{Z}=0$ if $\gamma (k)$ is not an integer. In the case of the tachyon and the graviton external fields now considered, these kinds of divergences essentially come from the Dirichlet boundary conditions along the directions transverse to the D-brane, which produce the negative term $-\frac{1}{2}k_{\bot }^{\ 2}$ in $\gamma (k)$. \\
\indent
 Similar divergent factors ($\tilde{Z}^{k^2/2}$) may arise for closed-string external fields with higher masses ($k^2<0$) even in the absence of the D-brane and the $B$ field. Such factors, however, should cause no divergences for scattering amplitudes, since the amplitudes including an external line with a higher mass are obtained by taking the residue at the relevant pole of the (non-divergent) tree amplitude of the closed and open-string tachyons. For this reason we expect that the factor $\tilde{Z}^{\gamma (k)}$ does not cause any divergence in scattering amplitudes. Details will be left for future studies.



\newpage
\noindent
\begin{thebibliography}{[00]}
\bibitem{rf:1}	A. Sen, hep-th/9904207, J. High Energy Phys. {\bf 02}02 (2000), 016; hep-th/9805170, Int. J. Mod. Phys. {\bf A14}(1999),~4061; hep-th/9902105.\\
\bibitem{rf:2} P. Yi, Nucl. Phys. {\bf B550}(1999),~214; hep-th/9901169; \\
O. Bergman, K. Hori and P. Yi, Nucl. Phys. {\bf B580}(2000),~289; hep-th/0002223; \\
G. Gibbons, K. Hori and P. Yi, hep-th/0009061; \\
J. A. Minahan and B. Zwiebach, J. High Energy Phys. {\bf 0009}(2000), ~029; hep-th/0008231; \\
M. Kleban, A. Lawrence and S. Shenker, hep-th/0012081; \\ 
H. Kawai and T. Kuroki, hep-th/0106103;\\
I. Y. Park, hep-th/0106078.
\bibitem{rf:3}	C. Chu and P. Ho, Nucl. Phys. {\bf B550}(1999),~151; hep-th/9812219; \\
M. M. Sheikh-Jabbari,  Phys. ~Lett., {\bf B455} (1999) 129; hep-th/9901080; \\
F. Ardfaei and M. M. Sheikh-Jabbari, J. High Energy Phys. {\bf 02}(2000), ~016; hep-th/9810072; hep-th/9906161. \\
\bibitem{rf:4} N. Seiberg and E. Witten, J. High Energy Phys. {\bf 09}(1999), ~032, hep-th/9908142. \\
\bibitem{rf:5} J. A. Minahan and B. Zwiebach, hep-th/0008231, references there in. \\
\bibitem{rf:6}	A. Kokado, G. Konisi and T. Saito, Prog. Theor. Phys. {\bf 106} (2001), 645, hep-th/0103226 \\ Conclusions of this paper should be modified. Consequences of the NS string
in the text are valid also in the case of the bosonic string. \\
\bibitem{rf:7} G. Konisi and T. Saito, Prog. Theor. Phys. {\bf 55} (1976), 280. \\
T. Saito, Studia Humana et Naturalia (Kyoto Pref. Univ. of Medicine) {\bf 22} (1988), 63. \\
\bibitem{rf:8} M. Ademollo, A. D'adda, R. D'auria, E. Napolitano, P. Di Vecchia, F. Gliozzi and S. Sciuto, Nucl. Phys, B C{\bf 22}(1974), 189. \\
\bibitem{rf:9} T. Yoneya, Prog. Theor. Phys. {\bf 54} (1975), 526; \\
M. R. Garousi and C. Myers Nucl. Phys. {\bf B475} (1996) 193; hep-th/9603194;\\
A. Hashimoto and I. Klebanov, Nucl. Phys. Proc. Suppl. {\bf 55B} (1997) 118; hep-th/9611214. \\
\end{thebibliography}
\end{document}